\documentclass[12pt]{article}
\usepackage[latin1]{inputenc}
\usepackage[T1]{fontenc}
\usepackage{amsmath}
\usepackage{amsfonts}
\usepackage{amssymb}
\usepackage{color}
\usepackage{graphicx}
\usepackage[english]{babel}
\usepackage{lmodern}
\usepackage{hyperref}
\title{Induced Chern-Simons term by dimensional reduction}
\author{C.~D.~Fosco$^a$ and F.~A.~Schaposnik$^b$
\\
~
\\
~
\\
{\normalsize $^a\!$\it Centro At\'omico Bariloche and Instituto
Balseiro,}\\
{\normalsize $\!$\it Comisi\'on Nacional de Energ\'\i a At\'omica, 8400
Bariloche, Argentina}\\
{\normalsize $\!$\it }\\
{\normalsize $^b\!$\it Departamento de F\'\i sica, Universidad
Nacional de La Plata}\\ {\normalsize\it Instituto de F\'\i sica La Plata-CONICET}\\
{\normalsize\it C.C. 67, 1900 La Plata,
Argentina}
 }
\begin{document}
\date{\today}
\maketitle
\begin{abstract}
We derive an induced Abelian Chern-Simons (CS) term in \mbox{$2+1$}
dimensions, by dimensional reduction from  the finite-temperature theory of a
Dirac field with both vector and axial-vector couplings to two 
Abelian gauge fields, in $3+1$ dimensions.  
In our construction, the CS term emerges for the lowest Matsubara mode of the
vector Abelian field, by integrating the fermionic field, under the
assumption that the axial vector field is in a ``vacuum'' configuration. 
This configuration is characterized by a single number,
which in turn determines the coefficient of the induced CS term for the
Abelian vector field.
\end{abstract}
\section{Introduction}
 Quantum field theories in $2+1$ dimensions have some features that make
them an important subject of research, with great relevance both in
theoretical developments and phenomenological applications. Among the
latter, besides the celebrated Condensed Matter models involving planar
systems, we should also mention the dimensional reduction at
high-temperatures in some High Energy physics systems, typically Yang-Mills
theories~\cite{Appelquist:1981vg}, in the context of hot QCD.

Among the most characteristic properties of these theories, one of them
shows up when considering gauge invariant systems, since they allow for the
construction of a local, topological and gauge-invariant functional of the
gauge field, which breaks parity: the Chern-Simons (CS) term~\footnote{See,
for example~\cite{Dunne:1998qy} for a review.}. 
Unlike what happens in $3+1$ dimensions, parity is understood, in the
\mbox{$2+1$} dimensional context, to correspond to the reflection of just
one of the two spatial coordinates (changing both coordinates has unit
determinant: it is a rotation in $\pi$).

It has been realized some time ago that the CS term may appear in a system
as a relic of the integration of matter degrees of freedom which break
parity explicitly; indeed, this has been first realized when
evaluating the effective action for a massive Dirac field coupled to a
gauge field~\cite{DJT}, since the mass term in $2+1$ dimensions breaks parity.
One of the terms appearing in the effective action for the gauge field is
parity breaking, and becomes a CS term when the mass of the fermion tends to
infinity. It has moreover been realized that an explicit breaking is not
required, since a quantum breaking is unavoidable, leading to a properly 
called parity anomaly~\cite{Niemi:1983rq}, \cite{Redlich:1983kn,Redlich:1983dv}.
The induced CS term and related objects have been subsequently studied in many 
different contexts~\cite{Babu:1987rs} and from novel standpoints~\cite{Giombi:2011rz,Closset:2012vp,Gama:2015tuy,Bonora:2016ida}.

As originally pointed out in a well-honored work by
Deser, Jackiw, and Templeton~\cite{DJT}, one of the motivations to study  the
$d=3$ dimensional Chern-Simons (CS) action, is that it leads to a
topological mass term for the gauge field, which could possibly be
connected with the high-temperature limit of a $d=4$  quantum field theory.
This could result in a mass generation  for the resulting effective
Halmiltonian, as analyzed by S.~Weinberg~\cite{Wein}.
 
In a recent work R.~D.~Pisarski \cite{Pi} comes back to the possibility
that the topological  CS mass term may in fact provide the correct
infrared regulation at high temperatures exposing his doubts concerning
the possibility that a theta term  $\theta \tilde F F$ term in a $d = 4$
high-temperature gauge theory could be at the origin of such phenomenon.

In the present work we follow a different strategy to connect the
Chern-Simons action in $2+1$ spacetime dimensions by dimensional reduction from  a
finite-temperature $3+1$ dimensional theory, namely  Dirac field theory with vector and
axial vector couplings to two external Abelian gauge fields, in $3+1$
dimensions. 

The dimensional reduction is implemented here under two assumptions about the
gauge fields: the axial field is assumed to be in a vacuum state, while the
vector one belongs to the lowest, zero Matsubara frequency configuration. 

The structure of this paper is as follows: in Sect.~\ref{sec:thesys}, we
introduce a $3+1$ dimensional theory and the assumptions we make before
evaluating its effective action. Then, in~\ref{sec:imag}, we evaluate the
imaginary part: the induced CS term. 
The real part of the effective action is briefly discussed
in~\ref{sec:real}. In~\ref{sec:conc} we present our conclusions.
 
\section{The system}\label{sec:thesys}
We consider a massless Dirac field in $3+1$ dimensions, at a finite
temperature $T$, endowed with vector and axial-vector couplings to external
Abelian gauge fields $A_\mu$ and $B_\mu$, respectively. These fields will
be assumed to belong to some specific classes below, but we shall first introduce
them as if they were arbitrary, for the sake of clarity. With this in mind,
the Euclidean action ${\mathcal S}$ of the system, in the Matsubara
formalism, is given by the expression:
\begin{equation}\label{eq:defsf}
{\mathcal S}(\bar{\psi},\psi;A,B)\;=\; \int_0^\beta d\tau \, \int d^3x \,\bar{\psi}(\tau, x)
\big(\not \! \partial + i \not \!\! A + i \not \!\! B \gamma_5
\big)\psi(\tau, x) \;, 
\end{equation}
where $\tau$ is the Euclidean time, and we use conventions whereby
Boltzmann constant $k_B \equiv 1$, so that $\beta = \frac{1}{T}$.
Space-time coordinates shall be denoted by $x_\mu$, $\mu = 0,1,2,3$, such
that $x_0 \equiv \tau$, and $x \equiv (x_1,x_2,x_3)$ where $x_i$ ($i = 1,
2, 3$) are the spatial Cartesian coordinates~\footnote{Indices from the
middle of the Greek alphabet run over the same range as $\mu$, while those
from the middle of the Roman one correspond to spatial coordinates, and
have the same range as $i$.}.
On the other hand, Dirac's matrices $\gamma_\mu$ are Hermitian and satisfy
the relations: \mbox{$\{\gamma_\mu,\gamma_\nu\} = 2 \delta_{\mu\nu}$},
while $\gamma_5$  is given by $\gamma_5 = \gamma_0 \gamma_1 \gamma_2
\gamma_3 = \gamma_5^\dagger$.

We are interested in extracting the parity-breaking part of the effective
action $\Gamma(A,B)$ due to the Dirac field quantum fluctuations:
\begin{equation}\label{eq:defgab}
	e^{-\Gamma(A , B)} \;=\; \int {\mathcal D}\psi {\mathcal
	D}\bar{\psi} \, e^{- {\mathcal S}(\bar{\psi},\psi;A,B)} \;.
\end{equation}
In the Matsubara formalism, the fermionic fields are antiperiodic in the
imaginary time interval, namely,
\begin{equation}
	\psi(\tau + \beta, x) \;=\; - 	\psi(\tau, x)  \;\;,\;\;\;
	\bar{\psi}(\tau + \beta, x) \;=\; - \bar{\psi}(\tau, x)  \;,
\end{equation}
while bosonic ones, in particular the gauge fields, are periodic
\begin{equation}
	A_\mu(\tau + \beta, x) \;=\; A_\mu(\tau, x)  \;\;,\;\;\;
	B_\mu(\tau + \beta, x) \;=\; B_\mu(\tau, x)  \;.
\end{equation}
As a consequence, when considering the set of allowed vector and
axial-vector gauge transformations:
\begin{align}
& \psi(\tau, x) \; \to \; e^{- i \Omega_A(\tau,x)} \psi(\tau, x)
	\;,\;\; 
\bar{\psi}(\tau, x) \; \to \; e^{i \Omega_A(\tau,x)} \bar{\psi}(\tau, x)
\;,
\nonumber\\ 
& A_\mu(\tau, x)  \; \to \; A_\mu(\tau, x) + \partial_\mu \Omega_A(\tau,x)
\end{align}
and 
\begin{align}
& \psi(\tau, x) \; \to \; e^{- i \Omega_B(\tau,x) \gamma_5} \psi(\tau, x)
\; ,\;\; 
\bar{\psi}(\tau, x) \; \to \;  \bar{\psi}(\tau, x) 
e^{- i \Omega_B(\tau,x)\gamma_5} 
\nonumber\\ 
& B_\mu(\tau, x)  \; \to \; B_\mu(\tau, x) + \partial_\mu
	\Omega_B(\tau,x)\;,
\end{align}
respectively, the functions $\Omega_{A,B}$ must be required to satisfy:
\begin{equation}
\Omega_{A,B}(\beta,x) \;=\; \Omega_{A,B}(0,x) \,+\, 2 \pi n_{A,B}
\;, 
\end{equation}
where $n_A$ and $n_B$ are integers, which label the respective winding numbers
of the large gauge transformations.

Up to now, no restriction about the gauge-field configurations has been
implemented; let us now make them more explicit: firstly, since we have in
mind the high temperature regime, and the field is periodic, we can invoke
the usual decoupling of the lowest mode. Indeed, in the Matsubara Fourier
expansion of $A_\mu$:
\begin{equation}
A_{\mu}(\tau,x) = \beta^{-\frac{1}{2}} \, \sum_{n=-\infty}^{+\infty}
e^{\frac{2 n \pi i}{\beta}} A_{\mu}^{(n)}(x)
\end{equation}
we keep just the $n=0$ mode, since the remaining ones have masses which
increase with temperature:
$A_{\mu}(\tau,x) \sim A_{\mu}^{(n)}(x) \equiv A_{\mu}(x)$. Note that this
produces $4$ space-dependent components; we add the further constraint of
having a space-independent $A_0$. An alternative way of characterizing this
is to say that we only keep purely magnetic (static) field configurations: the
simplest non-trivial one allowing for the existence of a reduced (non-trivial)
effective action. 
Note that the time component of $A_\mu$, may be assumed to  depend on
$\tau$, what is gauge equivalent to a constant field.  

The axial field, on the other hand, is assumed to be a vacuum
configuration (vanishing electric and magnetic fields), since we are
using it just as a seed to produce parity
breaking in the reduced theory. As mentioned  in~\cite{Gross:1980br}, a
vacuum gauge field configuration corresponds to no spatial components and a
time dependent temporal component. This is consistent with assuming a
Maxwell action for that field, and looking for its lowest action
configuration.

Therefore, the class of configurations that we consider may be characterized as follows:
\begin{equation}
	\partial_j A_0 \;=\; 0 \;, \;\; \partial_j B_0 \;=\; 0 \;,\;\;\;
	\partial_0 A_j \;=\;0 \;\;, \;\;\; B_j \;=\;0
	\;\;\;  (j = 1,2,3).
\end{equation}

\section{Imaginary part of the effective action}\label{sec:imag}
In this Section, we shall evaluate the imaginary part of $\Gamma$, under
the previous assumptions about the gauge field configurations.

Furthermore, $A_0$ and $B_0$ (that can only depend on $\tau$) may be
rendered $\tau$-independent (constant), having, respectively, 
the values $\tilde{A}_0$ and $\tilde{B}_0$:
\begin{align}
A_0(\tau) &\to\; \tilde{A}_0 \,=\, \frac{1}{\beta} \int_0^\beta d\tau
	A_0(\tau) \nonumber\\
B_0(\tau) &\to\; \tilde{B}_0 \,=\,  \frac{1}{\beta} \int_0^\beta d\tau
	B_0(\tau)  \;,
\end{align}
by means of a gauge transformation of the
fermions:
\begin{align}
& \psi(\tau, x) \; \to \; 
	e^{ i \int_0^\tau d\tau' ( A_0(\tau') - \tilde{A}_0 )} 
	e^{ i  \gamma_5  \int_0^\tau d\tau' ( B_0(\tau') - \tilde{B}_0 )
	\gamma_5} \psi(\tau, x) \nonumber\\
	& \bar{\psi}(\tau, x) \; \to \;\bar{\psi}(\tau, x) 
	e^{- i \int_0^\tau d\tau' ( A_0(\tau') - \tilde{A}_0 )} 
	e^{ i \gamma_5 \int_0^\tau d\tau' ( B_0(\tau') - \tilde{B}_0 )} \;.
\end{align}
Note that this gauge transformation is ``small'', i.e., connected to the 
identity (its winding number vanishes).
Under this transformation, the axial part the gauge transformation
above does not produce a non trivial Jacobian. Indeed, 
denoting by $\Omega_B(\tau)$ the parameter of that transformation: 
\mbox{$\Omega_B(\tau) \equiv \int_0^\tau d\tau' ( B_0(\tau') - \tilde{B}_0 )$},
and
\begin{equation}
K_\mu = \frac{1}{4\pi^2} \epsilon_{\mu\nu\alpha\beta} A_\nu
\partial_\alpha A_\beta \;,
\end{equation}
we note that the anomalous Jacobian ${\mathcal J}$ is:
\begin{align}\label{eq:jacob}
	{\mathcal J} &=\; 
	e^{- i  \int_0^\beta d\tau \int d^3x \Omega_B(\tau) \partial_\mu
	K_\mu(\tau,x) } \;=\; 
	e^{- i \int_0^\beta d\tau \int d^3x  \Omega_B(\tau) 
	\partial_\tau K_0(x) } \,=\,e^{0} \,=\,1 \;,
\end{align}
where we used the property that, for the configurations we are dealing with, 
$K_j = 0$ ($j = 1,2,3$), plus the time independence of the spatial
components of $A_\mu$. The periodicity of $\Omega_B$ for a ``small''
transformation is implicitly assumed in the fact that the anomalous
Jacobian is known for transformations which do not change the boundary
conditions.

Therefore, we arrive to an equivalent (i.e., having identical effective
action) expression for the action:
\begin{equation}\label{eq:defsfeq}
	{\mathcal S}(\bar{\psi},\psi;A,B)= 
	\int_0^\beta d\tau \int d^3x \,\bar{\psi}(\tau, x)
	\big[\not \! \partial + i \gamma_j A_j(x) + i \gamma_0 ( \tilde{A}_0 + 
	\tilde{B_0} \gamma_5) \big]\psi(\tau, x) \;.
\end{equation}
In the expression above, the constant values of the temporal components of
the gauge fields can also be shifted by an integer number of $\frac{2
\pi}{\beta}$. 
On the other hand, had we wanted to completely decouple also the constant
fields $\tilde{A}_0$ and $\tilde{B}_0$ we should have performed a gauge
transformation which, in general, would have spoiled the boundary
conditions, namely, because $\frac{1}{2\pi} [\Omega_{A,B}(\beta,x) -
\Omega_{A,B} (0,x)] \notin {\mathbb Z}$.  

Parity is {\em explicitly\/}
broken by the presence of $B$, and since the imaginary part of $\Gamma$
coincides with its parity breaking part, one can obtain the former as the
odd part (under parity) of the effective action. Note that a non-explicit
(i.e., anomalous) breaking of parity cannot be obtained by this procedure,
which is adamant to $B_\mu$-independent contributions.  
To obtain the imaginary part of $\Gamma$, we begin from:
\begin{equation}
{\rm Im}\big[ \Gamma(A , B)\big] \;=\;\Gamma_{\rm odd}(A , B)
\;=\;
\;=\;\frac{1}{2} 
\int_{-\tilde{B}_0}^{+\tilde{B}_0}
d\tilde{B}_0 \, \frac{\partial}{\partial \tilde{B}_0} \,
\Gamma(A,B) \;,
\end{equation}
where $A$ and $B$ are implicitly assumed to belong to the class we are considering here, namely, $A_0 = \tilde{A}_0$, $A_j = A_j(x)$, $B_0 = \tilde{B}_0$, $B_j = B_j(x)$.

To proceed, we introduce an expansion in Matsubara modes for the fermions,
\begin{equation}
	\psi(\tau,x) \,=\, \beta^{-\frac{1}{2}} \,
	\sum_{n=-\infty}^{+\infty} \psi_n(x) e^{- i \omega_n \tau}
	\;,\;\;
	\bar{\psi}(\tau,x) \,=\, \beta^{-\frac{1}{2}} \,
	\sum_{n=-\infty}^{+\infty} \bar{\psi}_n(x) e^{ i \omega_n \tau}
	\;,
\end{equation}
obtaining an alternative form of the action with all the modes decoupled:
\begin{align}\label{eq:defsfeq1}
	{\mathcal S}(\bar{\psi},\psi;A,B) & = \; 
	\sum_{n=-\infty}^{+\infty} \int d^3x \,\bar{\psi}_n( x) \,{\mathcal D}_n
	\,	\psi_n (x) \;,\nonumber\\
{\mathcal D}_n & \equiv \; \not \!d  + i \gamma_0 (\omega_n + \tilde{A}_0 + \tilde{B_0} \gamma_5)  \;,
\end{align}
and $\not \!d$ denotes a Dirac operator in $3$ Euclidean dimensions:
\mbox{$\not \!d \equiv \gamma_j (\partial_j + i A_j(x))$},
but built with $4 \times 4$ Dirac matrices $\gamma_j$. 
We then see that:
\begin{align}\label{eq:gab1}
	e^{-\Gamma(A , B)} &=\; 
	{\rm det}\big[\not \! \partial + i \gamma_j A_j(x) 
	+ i \gamma_0 ( \tilde{A}_0 + \tilde{B_0} \gamma_5) \big] 
	\nonumber\\
&=\;\prod_{n=-\infty}^{+\infty} 
{\rm det}\big[ \not \!d  + i \gamma_0 (\omega_n + \tilde{A}_0 + 
	\tilde{B_0} \gamma_5) \big] \;,
\end{align}
and
\begin{equation}\label{eq:gab2}
	\Gamma(A , B) \;=\; - \sum_{n=-\infty}^{+\infty}
	{\rm Tr}{\rm log}\big[ \not \!d  + i \gamma_0 (\omega_n + \tilde{A}_0 + 
	\tilde{B_0} \gamma_5) \big] \;.
\end{equation}
Therefore,
\begin{equation}\label{eq:godd1}
	\Gamma_{\rm odd} (A , B) \;=\; - 
\frac{1}{2} \int_{-\tilde{B}_0}^{+\tilde{B}_0}
d\tilde{B}'_0 \, \sum_{n=-\infty}^{+\infty}
	{\rm Tr}\Big[ i \gamma_0 \gamma_5 
	\frac{1}{\not \!d  + i \gamma_0 (\omega_n + \tilde{A}_0 + 
	\tilde{B}'_0 \gamma_5)} \Big]
	\;,
\end{equation}
where ``${\rm Tr}$'' denotes trace over both spacetime arguments and Dirac
matrices' indices (the latter shall be denoted by ``${\rm tr}$''). 

We can produce a more explicit expression:
\begin{equation}\label{eq:godd2}
	\Gamma_{\rm odd} (A , B) \;=\; 
\frac{1}{2} \int_{-\tilde{B}_0}^{+\tilde{B}_0}
d\tilde{B}'_0 \, \sum_{n=-\infty}^{+\infty}\, {\mathcal Q}_n(A,B') \;,
\end{equation}
where
\begin{equation}
{\mathcal Q}_n(A,B) \;= \;- i \,
\int d^3x \, {\rm tr}\Big[ \gamma_0 \gamma_5 
\langle x | \frac{1}{\not \!d  + i \gamma_0 (\omega_n + \tilde{A}_0 + 
\tilde{B_0} \gamma_5)} |x\rangle \Big] \;,
\end{equation}
and we have adopted Dirac's bra-ket notation to denote operator kernels.

Let us note that, up to this point, we have not made any assumption about
the magnitude of the temperature. From now on, we shall assume that $T >>
|A_j|$, the spatial components of the gauge field.  The temporal components
of $A$ and $B$, on the other hand, are gauge-equivalent to the constants
$\tilde{A}_0$ and $\tilde{B}_0$ and therefore cannot be regarded as small
just by invoking a similar argument to the one used for the spatial
components.  However, the fact that they are constants allows us to treat
them exactly.  Indeed, we will expand in powers of $A_j$, since $\omega_n
>> A_j$, $\forall n$. The lowest non-vanishing contribution to ${\mathcal
Q}_n$ is of the second order in $A_j$, as it may be seen from the vanishing
of the Dirac traces for the previous two orders. 

Keeping just the second-order contribution, we see that
\begin{equation}\label{eq:qn1}
{\mathcal Q}_n(A,B)
\;=\; i \, \int d^3x \,{\rm tr}\Big[ \gamma_0 \gamma_5 
\langle x| G_n \,\gamma_j A_j \, 
G_n \, \gamma_k A_k \,  
G_n |x\rangle \Big] \;,
\end{equation}
where we have introduced the operator 
\begin{equation}
G_n \;=\; \frac{1}{\not \!d_0  + i \gamma_0 (\omega_n + \tilde{A}_0 + 
\tilde{B_0} \gamma_5)}
\;\;,\;\;\; 
\not \!d_0 \equiv \gamma_j \partial_j \;.
\end{equation}
Although one could use any representation for the Dirac's matrices, it is
rather convenient, in this calculation, to use the chiral representation,
built in terms of $\sigma_0 \equiv {\mathbb I}_{2\times 2}$ and the
standard Pauli's matrices $\sigma_j$:
\begin{equation}
\gamma_\mu \;=\; 
\left( 
\begin{array}{cc} 
0 & \sigma_\mu^\dagger \\
\sigma_\mu & 0 
\end{array}
\right) \;\;,\;\;\;
\gamma_5 \;=\; 
\left( 
\begin{array}{cc} 
\mathbb{I}_{2\times 2} & 0 \\
0 & - \mathbb{I}_{2\times 2} 
\end{array}
\right) \;\;.
\end{equation}
When used in (\ref{eq:qn2}), this leads to an equation which is naturally
decomposed into two terms, one for each chirality component, which in turn 
involve traces of $2 \times 2$ matrices:
\begin{equation}\label{eq:qn2}
{\mathcal Q}_n(A,B) \; =\; {\mathcal Q}^{L}_n(A,B) 
+ {\mathcal Q}^{R}_n(A,B) \
\end{equation}
\begin{align}
{\mathcal Q}^{L}_n(A,B) & =\;-\, \int d^3x \,
{\rm tr}\Big[ 
\langle x | (\not \! \nabla - \omega_n - \tilde{A_0} - \tilde{B}_0)^{-2} \,\gamma_j A_j \nonumber\\
& \times \; (\not \! \nabla - \omega_n - \tilde{A_0} - \tilde{B}_0)^{-1} \,\gamma_k A_k \, 
|x\rangle \Big] \nonumber\\
{\mathcal Q}^{R}_n(A,B) & =\;-\, \int d^3x \,
{\rm tr}\Big[ 
\langle x | (\not \! \nabla + \omega_n + \tilde{A_0} - \tilde{B}_0)^{-2} \,\gamma_j A_j \nonumber\\
& \times \; (\not \! \nabla + \omega_n + \tilde{A_0} - \tilde{B}_0)^{-1} \,\gamma_k A_k \, 
|x\rangle \Big] \;,
\end{align}
where $\not \! \nabla \equiv \sigma_j \partial_j$. 

After some algebra, we find that $\Gamma_{\rm odd}$ may be expressed as
\begin{align}\label{eq:godd3}
\Gamma_{\rm odd}(A,B) \;=\;- \frac{1}{4}\,
\,\sum_{n=-\infty}^{+\infty} \int d^3x \bigg\{
&{\rm tr}\Big[
\langle x | \big((\not \! \nabla - \omega_n - \tilde{A_0} -
	\tilde{B}_0)^{-1} \,\gamma_j A_j \big)^2 |x\rangle \Big]\nonumber\\
+ &{\rm tr}\Big[ \langle x | \big((\not \! \nabla + \omega_n + \tilde{A_0}
	- \tilde{B}_0)^{-1} \,\gamma_j A_j \big)^2 |x\rangle
	\Big]\nonumber\\
- & {\rm tr}\Big[ \langle x | \big((\not \! \nabla - \omega_n - \tilde{A_0}
	+ \tilde{B}_0)^{-1} \,\gamma_j A_j \big)^2 |x\rangle
	\Big]\nonumber\\
- & {\rm tr}\Big[ \langle x | \big((\not \! \nabla + \omega_n + \tilde{A_0}
	+ \tilde{B}_0)^{-1} \,\gamma_j A_j \big)^2 |x\rangle \Big] \bigg\} \;.
\end{align}
The structure of each term of the four terms inside the sum over $n$ is
identical, except for a global factor, to the one of the effective action
for a massive Dirac field in $2+1$ dimensions. The difference being the
values one should use for the respective masses, which shall 
depend on $n$, $\tilde{A}_0$, and $\tilde{B}_0$. One sees that only the odd part in
the fermion mass (of each mode) is needed, after inserting the known $2+1$ dimensional 
result into (\ref{eq:godd3}).

Keeping just the leading terms in the corresponding ``masses'' (we shall
discuss the next to leading terms below), one gets:
\begin{equation}
\Gamma_{\rm odd}(A,B) \,=\, \frac{i}{8\pi} \,\xi(\tilde{A}_0,\tilde{B}_0) 
\int d^3x \epsilon_{jkl}
A_j(x) \partial_k A_l(x) \;,
\end{equation}
where:
\begin{equation}
\xi(\tilde{A}_0,\tilde{B}_0) \;=\; \sum_{n=-\infty}^{\infty} 
\Big(
\frac{\omega_n+\tilde{A}_0 - \tilde{B}_0}{|\omega_n+\tilde{A}_0 - \tilde{B}_0|}  
\,-\,
\frac{\omega_n+\tilde{A}_0 + \tilde{B}_0}{|\omega_n+\tilde{A}_0 + \tilde{B}_0|}
\Big) \;.
\end{equation}
Thus, the odd part of the $3+1$ dimensional effective action looks like a
Chern-Simons action with an effective coefficient:
the function $\xi$. This function is expressed as a series in Matsubara frequencies space. 
A convenient way to render it in more appealing form without spoiling its gauge transformation properties
is by using Poisson summation, or, in this context, Selberg's trace formula~\cite{selberg}.
In this case it amounts to replacing the series over frequencies by another one where each term
is (anti) Fourier transformed:
\begin{equation}
\xi(\tilde{A}_0,\tilde{B}_0) \;=\; - \frac{2}{\pi} 
\,
\sum_{k=-\infty}^{\infty} (-1)^k {\mathcal P}(\frac{1}{k}) \, 
{\rm sin}(k \beta \tilde{B}_0){\rm cos}(k \beta \tilde{A}_0) \;,
\end{equation}
where ${\mathcal P}$ denotes Cauchy's principal value. In the present
context its effect would be to get rid of a possible contribution from the 
$k = 0$ term; however, that term vanishes by itself. Besides, in
 the remaining terms the principal value prescription 
is irrelevant and can be removed.
Then:
\begin{align}
\xi(\tilde{A}_0,\tilde{B}_0) &=\;\frac{4}{\pi} 
\,
\sum_{k=1}^{\infty} (-1)^{k-1} \,
\frac{{\rm sin}(k \beta \tilde{B}_0) \, {\rm cos}(k \beta \tilde{A}_0)}{k}\nonumber\\
 &=\;\frac{2}{\pi} \Big\{
 {\rm Im}\big[ {\rm log}( 1 + e^{i \beta (\tilde{B}_0 - \tilde{A}_0)}) \big] \,+\, {\rm Im}\big[ {\rm log}( 1 + e^{i \beta (\tilde{B}_0 +
 \tilde{A}_0)})\big] \Big\} \nonumber\\
& =\;\frac{2}{\pi} \Big\{
 {\rm arctan}\big[ {\rm tan}(\frac{\beta \tilde{B}_0 - \beta \tilde{A}_0}{2}) \big] 
 \,+\,  
 {\rm arctan}\big[ {\rm tan}(\frac{\beta \tilde{B}_0 + \beta \tilde{A}_0}{2}) \big] 
 \Big\}
 \end{align}
i.e.,
\begin{equation}
\xi \;=\; \frac{2}{\pi} \,
\int_0^\beta d\tau B_0(\tau) \;.
\end{equation}
Recalling the origin of this result, from the imaginary part of logarithmic
functions, we see that under large gauge transformations with winding
number equal to, say, $n$, then  $\int_0^\beta d\tau B_0(\tau)$ will follow
that winding: it is an angular function.

Finally we have, for $\Gamma_{\rm odd}$:
\begin{equation}\label{eq:godd}
\Gamma_{\rm odd}(A,B) \,=\, \frac{i}{4\pi^2} \,	
\int_0^\beta d\tau B_0(\tau)
\,
\int d^3x \epsilon_{jkl} A_j(x) \partial_k A_l(x) \;,
\end{equation}
which, we recall, has been obtained as the leading term in a high temperature  expansion.

Note that the previous result may be written in terms of a Chern-Simons (C-S) action ${\mathcal S}_{CS}$, defined (in our conventions) by: 
\begin{equation}
{\mathcal S}_{CS}(A) \;\equiv\; \frac{1}{8\pi} \, 
\int d^3x \epsilon_{jkl} A_j(x) \partial_k A_l(x) \;,
\end{equation}
as:
\begin{equation}\label{eq:main}
\Gamma_{\rm odd}(A,B) \,=\, i \, \frac{2}{\pi} \,	
\int_0^\beta d\tau B_0(\tau)
\,{\mathcal S}_{CS}(A)\;.
\end{equation}
This is the main result of this paper, namely, that an induced CS term
emerges for the lowest (i.e. massless) Matsubara mode of $A$, in the high
temperature limit. The coefficient of that term is determined by a
parameter which labels the vacuum configurations of the $B$ field. In the
next Section, we propose a possible reason whereby a non-trivial value for
such parameter may naturally emerge.

The previous equation relates the content of the $B_0$-field configuration
to the coefficient multiplying the C-S action. We note that $\int_0^\beta
d\tau B_0(\tau)$ may be identified with $\frac{\pi}{2} N$, with $N$ being
the number of fermionic flavours, had the induced C-S proceeded from a
$2+1$ dimensional calculation.

We note that the next-to-leading term contribution to the odd part of the
effective action does contain an extra derivative of the spatial components
of the gauge field, so that it has the structure:
\begin{equation}
\Gamma_{\rm odd}^{\rm sub} \,=\, i \, \chi(\tilde{A}_0,\tilde{B}_0) 
\, {\mathcal S}_{PC}(A) 
\end{equation}
where ${\mathcal S}_{PC}(A) = \frac{1}{4} \int d^3x F_{jk}^2$, where 
$\chi(\tilde{A}_0,\tilde{B}_0)$ is an odd function under the a reflection 
in $\tilde{B}_0$. Thus, this subleading contribution is indeed odd (and imaginary) 
although it does not contribute to the induced C-S term, as it should be.

\section{Real part of the effective action}\label{sec:real}
The axial gauge field configuration has been assumed to be in a vacuum
configuration, from the point of view of its corresponding action. Note,
however, that the real part of its effective action,  for the same kind of 
configuration
as before, will receive quantum corrections. Besides, in the same limit as
the one used for the imaginary part, the space-dependent part of $A_j$ is
suppressed.
Thus, the real part  of the effective action may also be conveniently
obtained by first introducing the Matsubara modes for the fermions, as we 
did for the
imaginary part. Besides, it will be extensive, so it is convenient to
introduce  the (real part of the) free energy per unit volume, $f(A,B)$:
\begin{align}
{\rm Re} \big[ \Gamma(A,B) \big] & = \; - \frac{1}{2} 
\sum_{n=-\infty}^{+\infty}{\rm Tr} 
\log \big({\mathcal D}_n^\dagger {\mathcal D}_n \big) 
\nonumber\\
& = \; - V \; \beta \; f(A,B) 
\end{align}
with:
\begin{align}
f(A,B) \, =\, \beta^{-1} \, \sum_{n=-\infty}^{+\infty}
\int \frac{d^3k}{(2\pi)^3} \Big\{ &
\log\big[ k^2 + (\omega_n + \tilde{A}_0 + \tilde{B}_0)^2 \big]
\nonumber\\
+ \; 
& \log\big[ k^2 + (\omega_n + \tilde{A}_0 - \tilde{B}_0)^2 \big] \Big\}\;.
\end{align}

The sum over $n$ can be performed using standard complex variable techniques, leading to:
\begin{align}
f(A,B)  \; = \, - \frac{1}{\beta} \int \frac{d^3k}{(2\pi)^3} \Big\{ 
& \log\big[ {\rm cosh}(\beta k) + {\rm cos} (\beta \tilde{A}_0 + 
\beta\tilde{B}_0) 
\big]
\nonumber\\
+ & \log\big[ {\rm cosh}(\beta k) + {\rm cos} (\beta \tilde{A}_0 - 
\beta\tilde{B}_0) 
\big] \Big\}\;,
\end{align}
from which one can subtract its zero-temperature (i.e., vacuum) part 
in order to render it finite. Besides, looking for extrema with respect to
$\tilde{B}_0$, we find the necessary condition
\begin{align}
0 \; = \, \int \frac{d^3k}{(2\pi)^3} \Big[
& \frac{ {\rm sin} (\beta \tilde{A}_0 + \beta\tilde{B}_0)}{{\rm cosh}(\beta k) + {\rm cos} (\beta \tilde{A}_0 + \beta\tilde{B}_0)} 
\big]
\nonumber\\
& - \, \frac{ {\rm sin} (\beta \tilde{A}_0 - \beta\tilde{B}_0)}{{\rm cosh}(\beta k) + {\rm cos} (\beta \tilde{A}_0 - \beta\tilde{B}_0)} 
\big] 
\Big]\;,
\end{align}
which can be satisfied for $\int_0^\beta d\tau B_0(\tau) = \pi$ (mod $\pi$).
Coming back to the imaginary part in (\ref{eq:main}), this implies:
\begin{equation}
\Gamma_{\rm odd}(A,B) \,=\,2\,i \, {\mathcal S}_{CS}(A)\;.
\end{equation}
 We see that that corresponds to the standard result for the induced CS term in 
 $2+1$ dimensions for two $2$-component fields.

\section{Conclusions}\label{sec:conc}
We have evaluated the dimensionally reduced effective action due to a Dirac
field in $3+1$ dimensions, in the presence of vector and axial vector gauge
fields. Under some assumptions abut the system, namely, a vacuum
configuration for the axial one, and a purely magnetic one for the vector
field, an induced CS term appears for the latter.
The mechanism whereby this happens may be understood as due to an
unbalance, due to the axial field, between the number of fermionic
Matsubara modes having positive and negative masses.
Note that a related reduction mechanism has been used in~\cite{Kikukawa:1997qh} to 
formulate the overlap prescription for the Dirac operator for lattice fermions 
in an odd number of dimensions, although without an explicit breaking of parity 
by an external axial field. The parity anomalous contribution, which, as explained, we do not 
study here, was shown in~\cite{Kikukawa:1997qh} to correspond to a specific prescription 
for the phase of the Dirac operator. 

As a direction for future work, we note that it might be possible for alternative 
physical mechanisms to produce an induced CS term. Even at zero temperature, perfect conductor 
boundary conditions on the boundaries of a compact spatial coordinate~\cite{Edery:2009vr} lead to an 
interesting $2+1$ dimensional structure, while for the Dirac field one could try using
bag model~\cite{Chodos:1974pn} (or related) conditions.

\section*{Acknowledgments}
We acknowledge Prof.\ G.\ Semenoff (UBC, Canada) and A.\ Edery (BU, Canada) for 
their useful comments and suggestions.

This research was supported by ANPCyT, CONICET,  UNCuyo and UNLP.


\end{document}